\documentclass[aps,pra,showpacs,twocolumn]{revtex4-1}
\usepackage{amssymb}
\usepackage{amsmath}
\usepackage{graphicx}
\usepackage{epsfig}
\usepackage{subfigure}
\usepackage{color}

\begin{document}

\title{Higher order coherence as witness of exceptional point in Hermitian
bosonic Kitaev dimer}
\author{D. K. He}
\author{Z. Song}
\email{songtc@nankai.edu.cn}

\begin{abstract}
The non-analyticity induced by exceptional points (EPs) has manifestations
not only in non-Hermitian but also in Hermitian systems. In this work, we
focus on a minimal Hermitian bosonic Kitaev model to reveal the dynamical
demonstration of EPs in a Hermitian system. It is shown that the EPs
separate the parameter space into four regions, in which the systems are
characterized by different equivalent Hamiltonians, including the harmonic
oscillator, the inverted harmonic oscillator, and their respective
counterparts. We employ the second-order intensity correlation to
characterize a nonequilibrium quantum phase transition by calculating the
time evolution of a trivial initial state. The results indicate that the
concept of the EP can be detected in a small Hermitian bosonic system.
\end{abstract}
\affiliation{School of Physics, Nankai University, Tianjin 300071, China}
\maketitle

\section{Introduction}

In general, the concept of exceptional points (EPs) \cite%
{kato1966,berry2004,heiss2012}, that are the degeneracies of non-Hermitian
operators, is regarded as an exclusive feature of non-Hermitian systems.
However, the presence of EPs may have an impact on Hermitian systems,
particularly when a non-Hermitian matrix is embedded within them. This is
because the EPs always induce non-analyticity, which may manifest not only
in non-Hermitian systems but also in Hermitian systems. Previous studies {%
\cite{jin2010physics, jin2011partitioning,jin2011a,zhang2013self}} on
non-interacting systems have demonstrated a link between the EPs of an
effective non-Hermitian system on a finite lattice and the dynamic behavior
observed in infinite-size Hermitian systems. On the other hand, recent
research on the bosonic Kitaev model has garnered significant interest \cite%
{McDonald_PRX,wang2019non,flynn2020deconstructing,del2022non,wang2022quantum,bilitewski2023manipulating,ughrelidze2024interplay,slim2024optomechanical,busnaina2024quantum,hu2024bosonic}%
.\ Unlike its fermionic counterpart, the Hermitian bosonic Kitaev model
incorporates a non-Hermitian core matrix within its Nambu representation.
Importantly, the existence of a non-Hermitian core matrix does not require
the system to be infinitely large. We note that an infinite degree of
freedom is always involved in a finite-size bosonic Kitaev model. This
suggests that EPs are not exclusively associated with non-Hermitian systems,
but also with Hermitian systems that have an infinite degree of freedom. It
is our goal to identify correlated observables that demonstrate the
existence of EPs in a concrete Hermitian system.

In this work, we provide an alternative approach for bosonic Kitaev models.
We focus on a Hermitian bosonic Kitaev dimer to reveal the dynamical
demonstration of EPs using specific observables. Although the size of the
system is small, the EPs divide the parameter space into four regions. In
each region, the original Hamiltonian has different combinations of
equivalent Hamiltonians, including the harmonic oscillator (HO) and the
inverted harmonic oscillator (IHO). It is well known that the equivalent
Hamiltonians each have distinct dynamic characteristics, which can be used
to demonstrate the existence of EPs and identify non-equilibrium quantum
phase transitions. We employ the second-order intensity correlation to
characterize such transitions by calculating the time evolution of a trivial
initial state. The results indicate that the concept of the EP can be
detected in a small Hermitian bosonic system. The present scheme for
detecting QPTs has the following features. First, the quantity of
measurement is not a local observable as usual, but rather correlation
functions in the time domain. This is similar to the scheme used to witness
a QPT in the quantum Ising chain via entanglement \cite{Osborne T J,Vidal G}%
. Second, this transition is detected through quench dynamics, rather than
through measurements at zero temperature.

The structure of this paper is as follows. In Sec. \ref{Model and
exceptional points}, we introduce the model and pointed out the hidden
exceptional point within it. In Sec. \ref{Phase diagram}, we solve the
Hamiltonian exactly and present the phase diagram of the model. In Sec. \ref%
{Second-order coherence functions}, we introduce the second-order coherence
function to characterize the phase transition. Finally, in Sec. \ref{Summary}%
, we provide a summary and discussion.

\begin{figure}[ht]
\centering
\includegraphics[width=0.5\textwidth]{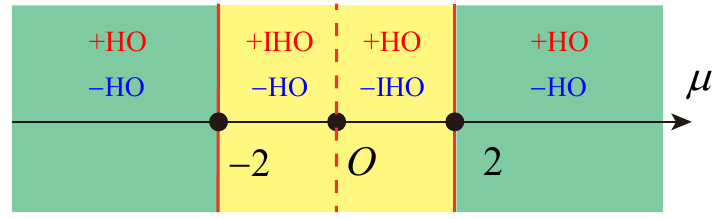}
\caption{Phase diagram of the Hamiltonian in Eq. (\protect\ref{H dimer}) on
the parameter $\protect\mu$ line. Different colors distinguish different
phases, and the distinction between phases lies in the different forms of
Eqs. (\protect\ref{H mu>2}) and (\protect\ref{H mu<2}). Overall, green
indicates that both $H^{+}$ and $H^{-}$ are HO, yellow indicates that one of 
$H^{+}$ and $H^{-}$ is a HO and the other is an IHO, and phase boundaries
are marked by red solid (or dashed) lines.}
\label{fig1}
\end{figure}

\section{Model and exceptional points}

\label{Model and exceptional points}

We consider a two-site Bose-Kitaev model with the Hamiltonian

\begin{equation}
H=\left( a^{\dag }+a\right) \left( b^{\dag }+b\right) +\mu \left( a^{\dag
}a+b^{\dag }b\right) ,  \label{H dimer}
\end{equation}%
where $a^{\dag }$ and $b^{\dag }$ $\left( a,b\right) $ are bosonic creation
(annihilation) operators. Here, $\mu $ represents the chemical potential
intensity for both types of bosons, and the strengths of both the hopping
and pairing terms are set to unity. This model can be regarded as the
effective Hamiltonian of the Dicke model, following the Holstein-Primakoff
transformation in the large-atom-number limit \cite{HePRB}. Intuitively, it
is the simplest Bose-Kitaev model, which is apparently trivial.
Nevertheless, unlike its fermionic counterpart, this particular bosonic
Kitaev dimer involves an infinite number of degrees of freedom, which may
support a continuous quantum phase transition. In fact, the Hilbert space is
spanned by a basis set $\left\{ \left\vert m,n\right\rangle ,m,n\in \lbrack
0,\infty )\right\} $\ of infinite dimensions, where the Fock basis is%
\begin{equation}
\left\vert m,n\right\rangle =\frac{1}{\sqrt{m!n!}}\left( a^{\dag }\right)
^{m}\left( b^{\dag }\right) ^{n}\left\vert 0\right\rangle _{a}\left\vert
0\right\rangle _{b},
\end{equation}%
and kets $\left\vert 0\right\rangle _{a}$\ and $\left\vert 0\right\rangle
_{b}$\ are the vacuum states of the two types of boson operators.

To proceed, we start to investigate the solution of the Hamiltonian.
Introducing a linear transformation%
\begin{equation}
d_{\pm }=\frac{1}{\sqrt{2}}\left( a\pm b\right) ,
\end{equation}%
the Hamiltonian can be expressed in a block-diagonalized form%
\begin{eqnarray}
H &=&H^{+}+H^{-}  \notag \\
&=&\varphi _{L}\left( 
\begin{array}{cc}
h^{+} & 0 \\ 
0 & h^{-}%
\end{array}%
\right) \varphi _{R},
\end{eqnarray}%
where the operator vectors are defined as $\varphi _{\mathrm{L}}=\left(
d_{+},-d_{+}^{\dagger },d_{-},-d_{-}^{\dagger }\right) $ and $\varphi _{%
\mathrm{R}}=\left( d_{+}^{\dagger },d_{+},d_{-}^{\dagger },d_{-}\right) ^{T}$%
. We note that both two matrices $h^{\pm }$, given by%
\begin{eqnarray}
h^{\pm } &=&\frac{\mu }{2}\left( 
\begin{array}{cc}
1 & 0 \\ 
0 & -1%
\end{array}%
\right) \pm \frac{1}{2}\left( 
\begin{array}{cc}
1 & 1 \\ 
-1 & -1%
\end{array}%
\right)   \notag \\
&=&\frac{\left( \mu \pm 1\right) }{2}\sigma _{z}\pm \frac{i}{2}\sigma _{y},
\end{eqnarray}%
are non-Hermitian, where $\sigma _{z}$ and $\sigma _{y}$ are Pauli matrices,
defined as%
\begin{equation}
\sigma _{z}=\left( 
\begin{array}{cc}
1 & 0 \\ 
0 & -1%
\end{array}%
\right) ,\sigma _{y}=\left( 
\begin{array}{cc}
0 & -i \\ 
i & 0%
\end{array}%
\right) .
\end{equation}%
It is easy to check that matrix $h^{-}$\ has an EP at $\mu =0$ and $2$,
while matrix $h^{+}$\ has an EP at $\mu =0$ and $-2$. In fact, the
eigenvalues of matrices $h^{\pm }$ can be expressed in the form 
\begin{eqnarray}
\lambda _{1}^{\pm } &=&-\frac{1}{2}\sqrt{\mu ^{2}\pm 2\mu },  \notag \\
\lambda _{2}^{\pm } &=&+\frac{1}{2}\sqrt{\mu ^{2}\pm 2\mu }.
\end{eqnarray}%
The corresponding eigenvectors are%
\begin{equation}
\phi _{1}^{\pm }=\left( 
\begin{array}{c}
\mp \lambda _{1}^{\pm }\mp \mu -1 \\ 
1%
\end{array}%
\right) ,\phi _{2}^{\pm }=\left( 
\begin{array}{c}
\mp \lambda _{2}^{\pm }\mp \mu -1 \\ 
1%
\end{array}%
\right) ,!!
\end{equation}%
satisfying the equations $h^{\pm }\phi _{1}^{\pm }=\lambda _{1}^{\pm }\phi
_{1}^{\pm }$ and $h^{\pm }\phi _{2}^{\pm }=\lambda _{2}^{\pm }\phi _{2}^{\pm
}!!$. We find that these eigenvalues experience non-analytical points at $%
\mu =0$ and $\pm 2$, at which the eigenvetors with zero eigenvalues coalesce.

In general, a finite-dimensional Hermitian matrix always has analytical
eigenvalues as the matrix elements vary. Technically, this is the reason why
a QPT cannot occur in a Hermitian system with a finite degree of freedom. In
contrast, a finite-dimensional non-Hermitian matrix can have non-analytical
eigenvalues at an EP as the matrix elements vary.\ For the present dimer, on
the one hand, it is Hermitian and has an infinite degree of freedom; on the
other hand, its core matrix is non-Hermitian and has EPs. These motivate us
to investigate such a dimer system and aim for the connection between the
EPs and the possible QPT of the system.

\section{Phase diagram}

\label{Phase diagram}

In this section, we will demonstrate that the EPs at $\mu =0$\ and $\pm 2$\
are the phase boundaries, at which the eigenstates become non-analytical.
Before proceeding with the solutions in each region, we would like to draw a
simple physical picture to address the main points. We consider a simplified
version of the Hamiltonian, which reads%
\begin{equation}
H_{0}=a^{\dag }b^{\dag }+ab+\mu \left( a^{\dag }a+b^{\dag }b\right) ,
\end{equation}%
in which the transition term between two modes of bosons. In this situation,
there are many invariant subspaces for $H_{0}$ arising from the symmetries%
\begin{equation}
\left[ \Pi ,H_{0}\right] =0,
\end{equation}%
where $\Pi $\ is the boson number parity operator%
\begin{equation}
\Pi =(-1)^{a^{\dag }a+b^{\dag }b}.
\end{equation}%
Here, we consider the case of a simply invariant subspace, which is spanned
by a basis set $\left\{ \left\vert l\right\rangle ,l\in \lbrack 0,\infty
)\right\} $%
\begin{equation}
\left\vert l\right\rangle =\frac{1}{l!}\left( a^{\dag }b^{\dag }\right)
^{l}\left\vert 0\right\rangle _{a}\left\vert 0\right\rangle _{b},
\end{equation}%
with $\Pi \left\vert l\right\rangle =\left\vert l\right\rangle $. Based on
this set of basis, the equivalent Hamiltonian of $H_{0}$\ can be expressed
as the form%
\begin{equation}
H_{0}^{\mathrm{eq}}=\sum_{l=0}^{\infty }[(\left( l+1\right) \left\vert
l+1\right\rangle \left\langle l\right\vert -\text{H.c.})+2\mu l\left\vert
l\right\rangle \left\langle l\right\vert ],!!!  \label{H_k_eq}
\end{equation}%
which describes a single-particle chain with an $l$-dependent
nearest-neighbor (NN) hopping term and linear potential. Although we still
have the solutions of the equivalent Hamiltonian, it provides us with an
intuitive physical picture of two distinct dynamic behaviors. We consider
the time evolution of the initial state $\left\vert 0\right\rangle
_{a}\left\vert 0\right\rangle _{b}$ under the systems with two extreme
values of the chemical potentials, $\mu =0$ and $\infty $, respectively. The
dynamics are equivalent to the relaxation process for a single particle
initially set at the end of a semi-infinite chain. In this sense, the
profile of the evolved state is straightforward in the two cases. (i) In the
case where $\mu =0$, the single particle should spread out over time, and
the probability never revives at the end of the chain. (ii) In the case
where $\mu =\infty $, the tunneling from the state $\left\vert
0\right\rangle _{a}\left\vert 0\right\rangle _{b}$\ to $\left\vert
1\right\rangle _{a}\left\vert 1\right\rangle _{b}$ is forbidden. This simple
analysis indicates that these two dynamic behaviors are completely
different. In the following sections, we will provide more detailed
investigations.

We start with the solution of the Hamiltonian to reveal its non-analyticity.
Although explicit solutions in each regions are not completely provided, we
will show that the solutions in each regions belong to different types of
equivalent Hamiltonians and then cannot cross over with each other as the
parameter passes the EP.

We will focus on the Hamiltonian $H$ for the case $\mu >0$ and extend the
results to the case $\mu <0$ using the relation $H\left( \mu ,b,b^{\dag
}\right) =-H\left( -\mu ,-b,-b^{\dag }\right) $. It turns out that the core
matrix is undiagonable at EPs. There is a second-order EP\ at $\left\vert
\mu \right\vert =2$, and two second-order EPs\ at $\mu =0$. It is presumably
the case that such nonanalyticity should result in the nonanalytic behaviors
of the eigenstates and the spectrum of the Hamiltonian $H$. In the
following, we will demonstrate this point by rewriting the Hamiltonian in
the analytical regions $0<\mu <2$ and $\mu >2$, respectively. The phase
diagram is schematically illustrated in Fig. \ref{fig1}.

(i) In the region $\mu >2$, we introduce a pair of Bogoliubov modes defined
as%
\begin{equation}
\gamma _{\rho }=\sinh \frac{\theta _{\rho }}{2}d_{\rho }^{\dag }+\cosh \frac{%
\theta _{\rho }}{2}d_{\rho },
\end{equation}%
with $\rho =\pm $, where $\tanh \left( \theta _{\rho }/2\right) =1-\rho
\left( \mu +\sqrt{\mu ^{2}+2\rho }\right) $, satisfying the canonical
relations%
\begin{eqnarray}
\left[ \gamma _{\rho },\gamma _{\rho ^{\prime }}^{\dag }\right] &=&\delta
_{\rho \rho ^{\prime }},  \notag \\
\left[ \gamma _{\rho },\gamma _{\rho ^{\prime }}\right] &=&\left[ \gamma
_{\rho }^{\dag },\gamma _{\rho ^{\prime }}^{\dag }\right] =0.
\end{eqnarray}%
The Hamiltonian can be written as the form $H=H^{+}+H^{-}$, where%
\begin{eqnarray}
H^{+} &=&\frac{1}{2}\sqrt{\mu ^{2}+2\mu }\left( \gamma _{+}^{\dag }\gamma
_{+}+\frac{1}{2}\right) ,  \notag \\
H^{-} &=&\frac{1}{2}\sqrt{\mu ^{2}-2\mu }\left( \gamma _{-}^{\dag }\gamma
_{-}+\frac{1}{2}\right) ,  \label{H mu>2}
\end{eqnarray}%
satisfying the relation $\left[ H^{+},H^{-}\right] =0$. Obviously, it
represents two independent HO systems for bosons with two different
frequencies. The ground state is the common vacuum state $\left\vert \text{%
Vac}\right\rangle $ of the bosonic operators $\gamma _{+}$\ and $\gamma _{-}$%
, that is $\gamma _{\pm }\left\vert \text{Vac}\right\rangle =0$.{\ }The
dynamics of such a system is straightforward. Any given initial state
undergoes periodic time evolution with two frequencies: $\frac{1}{2}\sqrt{%
\mu ^{2}+2\mu }$\ and $\frac{1}{2}\sqrt{\mu ^{2}-2\mu }$.

(ii) In the region $0<\mu <2$, for $H^{-}$ we introduce another bosonic
operators, given by%
\begin{equation}
A_{-}=\sinh \frac{\varphi _{-}}{2}d_{-}^{\dag }+\cosh \frac{\varphi _{-}}{2}%
d_{-},
\end{equation}%
where $\tanh \left( \varphi _{-}/2\right) =[-1+\sqrt{-\left( \mu ^{2}-2\mu
\right) }]/\left( \mu -1\right) $, satisfying the canonical relations%
\begin{eqnarray}
\left[ A_{\rho },A_{\rho ^{\prime }}^{\dag }\right] &=&\delta _{\rho \rho
^{\prime }},  \notag \\
\left[ A_{\rho },A_{\rho ^{\prime }}\right] &=&\left[ A_{\rho }^{\dag
},A_{\rho ^{\prime }}^{\dag }\right] =0.
\end{eqnarray}%
The Hamiltonian can be written as the form $H=H^{+}+H^{-}$, where%
\begin{eqnarray}
H^{+} &=&\frac{1}{2}\sqrt{\mu ^{2}+2\mu }\left( \gamma _{+}^{\dag }\gamma
_{+}+\frac{1}{2}\right),  \notag \\
H^{-} &=&-\frac{1}{2}\sqrt{-\left( \mu ^{2}-2\mu \right) }\left[ \left(
A_{-}^{\dag }\right) ^{2}+\text{\textrm{h.c.}}\right],  \label{H mu<2}
\end{eqnarray}%
satisfying the relation $\left[ H^{+},H^{-}\right] =0$. It indicates that
the original system consists of two independent equivalent sub-systems. $%
H^{+}$ represents a HO system, while $H^{-}$ represents an IHO, which has
been shown to possess a continuous and unbounded energy spectrum. In
comparison with the ground state of the HO, the common vacuum state $%
\left\vert \text{Vac}\right\rangle $\ of the bosonic operators $A_{+}$\ and $%
A_{-}$, which satisfies $A_{\pm }\left\vert \text{Vac}\right\rangle =0$, is
no longer the ground state of the system. In particular, the time evolution
operator of an IHO Hamiltonian corresponds to a squeezing operator \cite%
{IHO1,IHO2,QO}. The squeezing operator is a fundamental concept in quantum
optics and quantum mechanics, used to describe the transformation of quantum
states, particularly in the context of squeezed states. This allows us to
characterize the dynamics in a rigorous manner. In fact, considering the
common vacuum state $\left\vert \text{Vac}\right\rangle $ as the initial
state,\ the expectation value of the particle number for the evolved state
can be obtained as%
\begin{equation}
\left\langle A_{-}^{\dag }A_{-}\right\rangle =\frac{1}{\left\vert \tanh 
\sqrt{2\mu -\mu ^{2}}t\right\vert ^{-2}-1},
\end{equation}%
which is obviously no loner periodic, but increases exponentially.
Importantly, such a dynamic behavior should mask the periodic dynamics
arising from the sub-Hamiltonian $H^{+}$. This is crucial to distinguish the
two regions in the phase diagram from the perspective of dynamics.
\begin{figure*}[ht]
\centering
\includegraphics[width=1\textwidth]{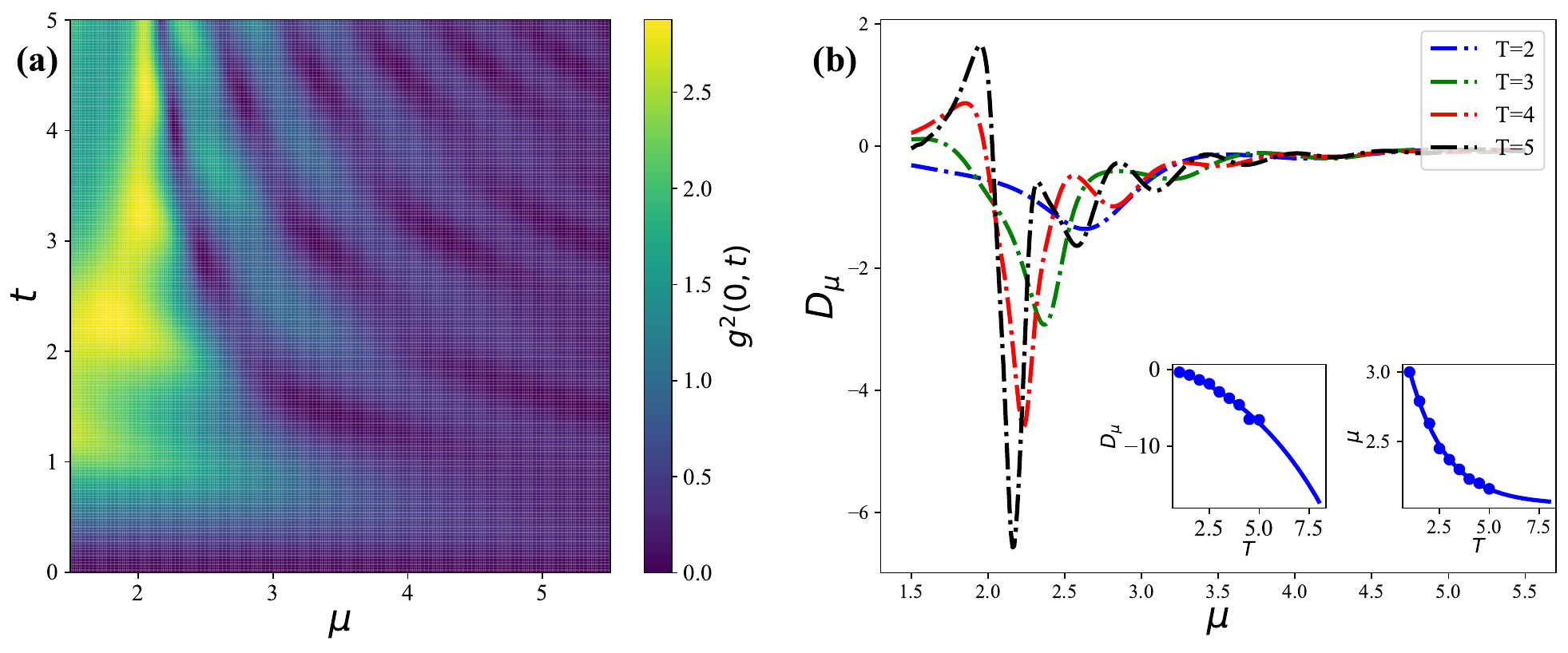}
\caption{Plots of the second-order coherence function $g^{(2)}(0,t)$ and the
quantity $D_{\protect\mu }$, given by Eq. (\protect\ref{second order}) and
Eq.(\protect\ref{Dm}). The numerical simulation is performed by tracking the
time evolution of the initial state Eq. (\protect\ref{is}). The quantity $%
g^{(2)}(0,t)$ for a given $\protect\mu $ is obtained by exact
diagonalization of the finite-dimensional matrix representation of
Hamiltonian. (a) A three-dimensional (3D) plot of $g^{(2)}(0,t)$ as a
function of time $t$ and $\protect\mu $. (b) Plots of $D_{\protect\mu }$,
the derivative second-order coherence function $g^{(2)}(0,t)$ over a period
of time $T$ with respect to $\protect\mu $, as defined by Eq. (\protect\ref%
{Dm}). The time interval $T$ are indicated in the legend. The results show
that each $D_{\protect\mu }$ has a minimum near the EP at $\protect\mu =2$.
The valleys become deeper as $T$ increases, implying that $D_{\protect\mu }$
is divergent at the EP in the case of infinite $T$. Inset: The minimum value
of $D_{\protect\mu }$ and the corresponding value of $\protect\mu $ for
different truncation times $T$. The blue circles represent the scatter plots
of minimum $D_{\protect\mu }$ and the corresponding value of $\protect\mu $
for Eq. (39), respectively, while the blue dashed lines are their respective
fitting curves.}
\label{fig2}
\end{figure*}
Then, we conclude that the non-analyticity arising from the EP at $\mu =2$
can be demonstrated through the quench dynamics. This finding is in
accordance with the nonequilibrium phase transition in the Dicke model, as
proposed in Ref. \cite{Dicke_IHO}. These results can be directly applied to
the case with $\mu <0$. In fact, the equivalent Hamiltonians becomes HO for $%
\mu <-2$ and the composition of IHO and HO for $-2<\mu <0$, respectively.
The corresponding operators in the Hamiltonians can be obtained by taking\ $%
d_{\pm }\rightarrow d_{\mp }$. Here, we want to emphasize that the
fundamental reason for the non-periodic dynamical behavior of the
Hamiltonian $H$ in the $-2<\mu <0$ region is due to $H^{+}$. The phase
diagram of the $H$ is the result of the composition of $H^{+}$ and $H^{-}$.

\begin{figure*}[tbh]
\centering
\includegraphics[width=1\textwidth]{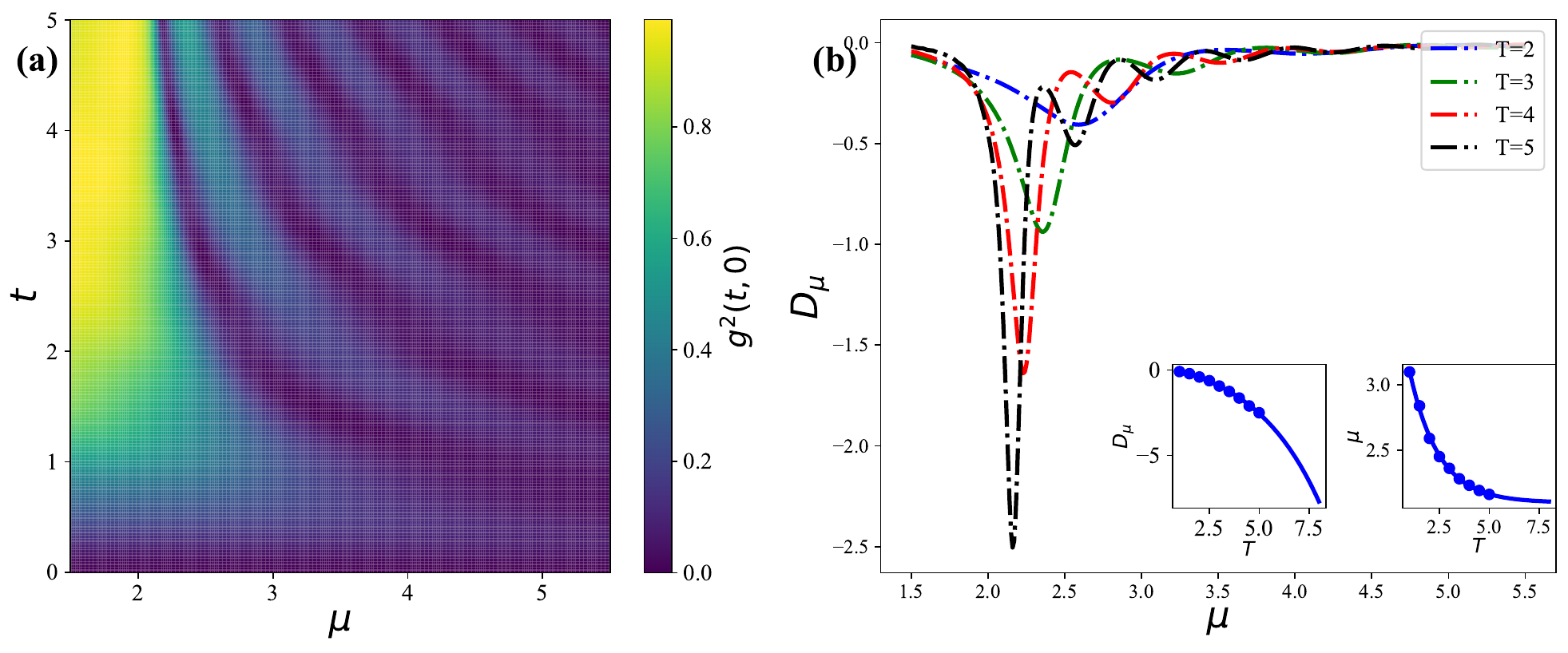}
\caption{Plots of the second-order coherence function $g^{(2)}(t,0)$ and the
quantity $D_{\protect\mu }$, given by Eq. (\protect\ref{second order}) and
Eq.(\protect\ref{Dm}). The numerical simulation is performed by tracking the
time evolution of the initial state Eq. (\protect\ref{is}). The quantity $%
g^{(2)}(0,t)$ for a given $\protect\mu $ is obtained by exact
diagonalization of the finite-dimensional matrix representation of
Hamiltonian. (a) A three-dimensional (3D) plot of $g^{(2)}(t,0)$ as a
function of time $t$ and $\protect\mu $. (b) Plots of $D_{\protect\mu }$,
the derivative second-order coherence function $g^{(2)}(0,t)$ over a period
of time $T$ with respect to $\protect\mu $, as defined by Eq. (\protect\ref%
{Dm}). The time interval $T$ are indicated in the legend. The results show
that each $D_{\protect\mu }$ has a minimum near the EP at $\protect\mu =2$.
The valleys become deeper as $T$ increases, implying that $D_{\protect\mu }$
is divergent at the EP in the case of infinite $T$. Inset: The minimum value
of $D_{\protect\mu }$ and the corresponding value of $\protect\mu $ for
different truncation times $T$. The blue circles represent the scatter plots
of minimum $D_{\protect\mu }$ and the corresponding value of $\protect\mu $
for Eq. (39), respectively, while the blue dashed lines are their respective
fitting curves.}
\label{fig3}
\end{figure*}

\section{Second-order coherence functions}

\label{Second-order coherence functions}

In this section, we focus on how to measure the different dynamic behaviors
in each region of the phase diagram. In our previous work \cite{HePRB}, we
have employed the expectation value of the particle number operator $a^{\dag
}a$ for the state evolved from the vacuum state to detect the phase
transition of the system. In the following, we will demonstrate that the
second-order coherence function can serve as a witness to measure the phase
transition of the Hamiltonian.

For a given initial state $\left\vert \psi \left( 0\right) \right\rangle $,
its time evolution at instant $t$\ is

\begin{equation}
\left\vert \psi \left( t\right) \right\rangle =\exp \left( -iHt\right)
\left\vert \psi \left( 0\right) \right\rangle .
\end{equation}%
We introduce the second-order coherence function, given by 
\begin{widetext}
	\begin{equation}
		g^{\left( 2\right) }\left( t_{1},t_{2}\right) =\frac{\left\langle \psi
			\left( t_{2}\right) |a^{\dag }\left( 0\right) a^{\dag }\left( t_{1}\right)
			a\left( t_{1}\right) a\left( 0\right) |\psi \left( t_{2}\right)
			\right\rangle }{\left\langle \psi \left( t_{2}\right) |a^{\dag }\left(
			t_{1}\right) a\left( t_{1}\right) |\psi \left( t_{2}\right) \right\rangle
			\left\langle \psi \left( t_{2}\right) |a^{\dag }\left( 0\right) a\left(
			0\right) |\psi \left( t_{2}\right) \right\rangle }, \label{second order}
	\end{equation}
\end{widetext}{where the operator }$a${\ in the Heisenberg picture is $%
a\left( t\right) =\exp \left( iHt\right) a\exp \left( -iHt\right) $.} This
function is used to differentiate between optics experiments that require a
quantum mechanical description and those for which classical fields are
sufficient. In comparison to the time evolution of the average particle
number, the first-order coherence function deals with amplitude
correlations, while the second-order coherence deals with intensity
correlations. In the following, the numerical simulations for two quantities 
$g^{\left( 2\right) }\left( 0,t\right) $\ and $g^{\left( 2\right) }\left(
t,0\right) $\ are performed for a simple initial state

\begin{equation}
\left\vert \psi \left( 0\right) \right\rangle =\frac{1}{\sqrt{2}}\left(
a^{\dag }+b^{\dag }\right) \left\vert 0\right\rangle .  \label{is}
\end{equation}%
We expect that the two quantities $g^{\left( 2\right) }\left( 0,t\right) $\
and $g^{\left( 2\right) }\left( t,0\right) $\ can be employed to measure the
EP, as signatures of the nonequilibrium quantum phase transitions.

The numerical simulation is performed by computing the time evolution of an
initial state $\left\vert \psi \left( 0\right) \right\rangle $ under the
original Hamiltonian $H$ given by the Eq. (\ref{H dimer}). The results of
two quantities $g^{\left( 2\right) }\left( 0,t\right) $\ and $g^{\left(
2\right) }\left( t,0\right) $ during the evolution process can be seen in
Figs. \ref{fig2} and \ref{fig3}. In order to detect the accurate transition
point, we also calulate the derivative of the average $g^{\left( 2\right)
}\left( 0,t\right) $\ or $g^{\left( 2\right) }\left( t,0\right) $ over a
period of time $T$ with respect to $\mu $,

\begin{equation}
D_{\mu }=\frac{\partial }{\partial \mu }\left[ \frac{1}{T}%
\int_{0}^{T}g^{\left( 2\right) }\text{\textrm{d}}t\right] .  \label{Dm}
\end{equation}%
These quantities are obtained by exact diagonalization of the
finite-dimensional matrix representation of original Hamiltonian. We observe
that the curves of $D_{\mu }$ indicate the quasi-critical points near $\mu
=\pm 2$. The corresponding extrapolation fitting shows that the quantities\ $%
D_{\mu }$ are divergent at the EP in the case of infinite $T$. The result
indicates that two quantities $g^{\left( 2\right) }\left( 0,t\right) $\ and $%
g^{\left( 2\right) }\left( t,0\right) $ are good candidates to witness the
EPs.

\section{Summary}

\label{Summary}

In summary, we have demonstrated the possible connection between
nonequilibrium quantum phase transitions in a Hermitian system and the EPs
hidden within it, based on the analysis of the equivalent Hamiltonians of
the bosonic Kitaev dimer. We have shown that, although this system has a
minimal size, the core matrix of its Nambu representation is non-Hermitian
due to the bosonic statistics. We have found that the nonanalyticity induced
by the EPs of the core matrix results in the equivalent Hamiltonian of the
original one being different in each analytical region. We also propose a
scheme for the experimental detection of the transition, based on the
measurement of the second-order intensity correlation for the quench
dynamics starting from the selected trivial state. Our work provides an
alternative way to demonstrate the existence of the EPs in a minimal
Hermitian system.

\section*{Acknowledgment}

We acknowledge the support of NSFC (Grants No. 12374461).


\begin{thebibliography}{99}
\bibitem{kato1966} Tosio Kato. 
\newblock {\em Perturbation theory for
linear operators}, volume 132. \newblock Springer Science \& Business Media,
2013.

\bibitem{berry2004} Michael~V Berry. \newblock Physics of nonhermitian
degeneracies. \newblock {\em Czechoslovak journal of physics},
54(10):1039--1047, 2004.

\bibitem{heiss2012} Walter~D Heiss. \newblock The physics of exceptional
points. \newblock {\em Journal of Physics A: Mathematical and Theoretical},
45(44):444016, 2012.

\bibitem{yang2018dynamical} XM~Yang, XZ~Zhang, C~Li, and Z~Song. \newblock %
Dynamical signature of the moir{\'e} pattern in a non-hermitian ladder. %
\newblock {\em Phys. Rev. B}, 98(8):085306, 2018.

\bibitem{zhang2020resonant} KL~Zhang and Z~Song. \newblock %
Resonant-amplified and invisible bragg scattering based on spin coalescing
modes. \newblock {\em Phys. Rev. B}, 102(10):104309, 2020.

\bibitem{zhang2019helical} KL~Zhang, L~Jin, and Z~Song. \newblock Helical
resonant transport and purified amplification at an exceptional point. %
\newblock {\em Phys. Rev. B}, 100(14):144301, 2019.

\bibitem{Doppler2016} Doppler, Alexei~A. Mailybaev, Julian B, Ulrich Kuhl,
Adrian Girschik, Florian Libisch, Thomas~J. Milburn, Peter Rabl, Nimrod
Moiseyev, and Stefan Rotter. \newblock Dynamically encircling an exceptional
point for asymmetric mode switching. \newblock {\em Nature},
537(7618):76--79, September 2016.

\bibitem{Xu2016} H.~Xu, D.~Mason, Luyao Jiang, and J.~G.~E. Harris. %
\newblock Topological energy transfer in an optomechanical system with
exceptional points. \newblock {\em Nature}, 537(7618):80--83, September 2016.

\bibitem{Assawaworrarit2017} Sid Assawaworrarit, Xiaofang Yu, and Shanhui
Fan. \newblock Robust wireless power transfer using a nonlinear
parity-time-symmetric circuit. \newblock {\em Nature}, 546(7658):387--390,
June 2017.

\bibitem{Wiersig2014} Jan Wiersig. \newblock Enhancing the sensitivity of
frequency and energy splitting detection by using exceptional points:
Application to microcavity sensors for single-particle detection. \newblock
\emph{Phys. Rev. L}, 112:203901, May 2014.

\bibitem{Wiersig2016} Jan Wiersig. \newblock Sensors operating at
exceptional points: General theory. \newblock {\em Phys. Rev. A}, 93:033809,
Mar 2016.

\bibitem{Hodaei2017} Hossein Hodaei, Absar~U. Hassan, Steffen Wittek,
Hipolito Garcia-Gracia, Ramy El-Ganainy, Demetrios~N. Christodoulides, and
Mercedeh Khajavikhan. \newblock Enhanced sensitivity at higher-order
exceptional points. \newblock {\em Nature}, 548(7666):187--191, August 2017.

\bibitem{Chen2017} Weijian Chen, ahin Kaya~zdemir, Guangming Zhao, Jan
Wiersig, and Lan Yang. \newblock Exceptional points enhance sensing in an
optical microcavity. \newblock {\em Nature}, 548(7666):192--196, August 2017.

\bibitem{jin2010physics} L~Jin and Z~Song. \newblock Physics counterpart of
the pt non-hermitian tight-binding chain. \newblock {\em Phys. Rev. A},
81(3):032109, 2010.

\bibitem{jin2011partitioning} L~Jin and Z~Song. \newblock Partitioning
technique for discrete quantum systems. \newblock {\em Phys. Rev. A},
83(6):062118, 2011.

\bibitem{jin2011a} L~Jin and Z~Song. \newblock A physical interpretation for
the non-hermitian hamiltonian. \newblock {\em Phys. Rev. A}, 44(37):375304,
2011.

\bibitem{zhang2013self} XZ~Zhang, L~Jin, and Z~Song. \newblock %
Self-sustained emission in semi-infinite non-hermitian systems at the
exceptional point. \newblock {\em Phys. Rev. A}, 87(4):042118, 2013.

\bibitem{McDonald_PRX} Alexander McDonald, T~Pereg-Barnea, and AA~Clerk. %
\newblock Phase-dependent chiral transport and effective non-hermitian
dynamics in a bosonic kitaev-majorana chain. \newblock {\em Phys. Rev. X},
8(4):041031, 2018.

\bibitem{wang2019non} Yu-Xin Wang and AA~Clerk. \newblock Non-hermitian
dynamics without dissipation in quantum systems. 
\newblock {\em Phys. Rev.
A}, 99(6):063834, 2019.

\bibitem{Osborne T J} Osborne T J, Nielsen M A. Entanglement in a simple
quantum phase transition[J]. \newblock {\em Phys. Rev. A}, 66(3): 032110,
2002.

\bibitem{Vidal G} Vidal G, Latorre J I, Rico E, et al. Entanglement in
quantum critical phenomena[J]. \newblock {\em Phys. Rev. L}, 90(22): 227902,
2003.

\bibitem{flynn2020deconstructing} Vincent~P Flynn, Emilio Cobanera, and
Lorenza Viola. \newblock Deconstructing effective non-hermitian dynamics in
quadratic bosonic hamiltonians. \newblock {\em New Journal of Physics},
22(8):083004, 2020.

\bibitem{del2022non} Javier Del~Pino, Jesse~J Slim, and Ewold Verhagen. %
\newblock Non-hermitian chiral phononics through optomechanically induced
squeezing. \newblock {\em Nature}, 606(7912):82--87, 2022.

\bibitem{wang2022quantum} Ya-Nan Wang, Wen-Long You, and Gaoyong Sun. %
\newblock Quantum criticality in interacting bosonic kitaev-hubbard models. %
\newblock {\em Phys. Rev. A}, 106(5):053315, 2022.

\bibitem{bilitewski2023manipulating} Thomas Bilitewski and Ana~Maria Rey. %
\newblock Manipulating growth and propagation of correlations in dipolar
multilayers: From pair production to bosonic kitaev models. 
\newblock {\em
Phys. Rev. L}, 131(5):053001, 2023.

\bibitem{ughrelidze2024interplay} Mariam Ughrelidze, Vincent~P Flynn, Emilio
Cobanera, and Lorenza Viola. \newblock Interplay of finite-and infinite-size
stability in quadratic bosonic lindbladians. \newblock {\em
Phys. Rev. A}, 110(3):032207, 2024.

\bibitem{slim2024optomechanical} Jesse~J Slim, Clara~C Wanjura, Matteo
Brunelli, Javier Del~Pino, Andreas Nunnenkamp, and Ewold Verhagen. \newblock %
Optomechanical realization of the bosonic kitaev chain. \newblock
\emph{Nature}, 627(8005):767--771, 2024.

\bibitem{busnaina2024quantum} Jamal~H Busnaina, Zheng Shi, Alexander
McDonald, Dmytro Dubyna, Ibrahim Nsanzineza, Jimmy~SC Hung, CW~Sandbo Chang,
Aashish~A Clerk, and Christopher~M Wilson. \newblock Quantum simulation of
the bosonic kitaev chain. \newblock {\em Nature
Communications}, 15(1):3065, 2024.

\bibitem{hu2024bosonic} Jia-Ming Hu, Bo~Wang, and Ze-Liang Xiang. \newblock %
Bosonic holes in quadratic bosonic systems. 
\newblock {\em arXiv
preprint arXiv:2408.01059}, 2024.

\bibitem{kitaev2001unpaired} A~Yu Kitaev. \newblock Unpaired majorana
fermions in quantum wires. \newblock {\em Physics-uspekhi}, 44(10S):131,
2001.

\bibitem{IHO1} Subramanyan V, Hegde S S, Vishveshwara S, et al. Physics of
the Inverted Harmonic Oscillator: From the lowest Landau level to event
horizons[J]. Annals of Physics, 2021, 435: 168470.

\bibitem{IHO2} Barton G. Quantum mechanics of the inverted oscillator
potential[J]. Annals of Physics, 1986, 166(2): 322-363.

\bibitem{QO} Scully M O, Zubairy M S. Quantum optics[M]. Cambridge
university press, 1997.

\bibitem{Dicke_IHO} Gietka K, Busch T. Inverted harmonic oscillator dynamics
of the nonequilibrium phase transition in the Dicke model[J]. Physical
Review E, 2021, 104(3): 034132.

\bibitem{HePRB} He D K, Song Z. Hidden exceptional point and the
localization-delocalization phase transition in the Hermitian bosonic Kitaev
model[J]. Physical Review B, 2025, 111(3): 035131.
\end{thebibliography}
\end{document}